\documentclass[]{aa}
\usepackage{graphicx}
\usepackage{txfonts}
\usepackage{booktabs}
\usepackage{multirow}

\usepackage{lineno}
\usepackage[table,xcdraw]{xcolor}
\usepackage{xspace}
\begin{document}

   \title{Impact chronology of leftover planetesimals}

   \author{R. Brasser\inst{1}\fnmsep\inst{2}\thanks{rbrasser@konkoly.hu}}
   \institute{Konkoly Observatory, HUN-REN CSFK, MTA Centre of Excellence; Konkoly Thege Miklos St. 15-17, H-1121 Budapest, Hungary  
    \and 
   Centre for Planetary Habitability (PHAB), Department of Geosciences, University of Oslo; Sem Saelands Vei 2A, N-0371 Oslo, Norway}
\authorrunning{Brasser}
   \date{}

% \abstract{}{}{}{}{} 
% 5 {} token are mandatory
\abstract
{After the formation of the Moon, the terrestrial planets were pummelled by impacts from planetesimals left over from terrestrial planet formation. Most lunar crater chronologies are fitted with an exponentially decreasing impact rate with an e-folding time of about 150~Myr. Dynamical simulations consisting of leftover planetesimals and the asteroid belt should reproduce this impact rate.}
{This work attempts to reproduce the impact rates set by modern crater chronologies using leftover planetesimals from three different dynamical models of terrestrial planet formation.}
{I ran dynamical simulations for 1 billion years using leftover planetesimals from the grand tack, depleted disc, and implantation models of terrestrial planet formation with the CPU version of the Gravitational ENcounters with GPU Acceleration (GENGA) N-body integrator. I fitted the cumulative impacts on the Earth and Mars as well as the fraction of remaining planetesimals using a function that is a sum of exponentials with different weighing factors and e-folding times.} 
{Most fits require three or four terms. The fitted timescales cluster around $\tau_1=10$~million years (Myr), $\tau_2=35$~Myr, $\tau_3=100$~Myr, and $\tau_4>200$~Myr. I attribute them to dynamical losses of planetesimals through different mechanisms: high-eccentricity Earth-crossers and the $\nu_6$ secular resonance, Earth-crossers, Mars-crossers, and objects leaking onto Mars-crossing orbits from beyond Mars. I placed a constraint on the initial population using the known Archean terrestrial spherule beds, and I conclude that the Archean impacts were mostly created by leftover planetesimals. The inferred mass in leftover planetesimals at the time of the Moon's formation was about 0.015 Earth masses.}
{The third time constant, $\tau_3$, is comparable to that of modern crater chronologies. As such, the crater chronologies are indicative of impacts by an ancient population of Mars-crossers. The initial perihelion distribution of the leftovers is a major factor in setting the rate of decline: to reproduce the current crater chronologies, the number of Earth-crossers at the time of the Moon's formation had to be at most half that of the Mars-crossers. These results together place constraints on dynamical models of terrestrial planet formation.}

\maketitle
%
%-------------------------------------------------------------------
\section{Introduction}
In traditional dynamical models, the terrestrial planets grew from a coagulation of planetesimals into protoplanets, which remained submerged in a swarm of planetesimals for millions of years. This system eventually evolved into a giant impact phase, wherein the protoplanets collided with the planetesimals and with each other, which ultimately led to the terrestrial planets. There exist many different dynamical simulations to explain this history \citep[e.g.][]{Chambers2001,Raymond2006,Raymond2009,OBrien2006,WL2019,Woo2021}; a review can be found in \citet{Morbidelli2012}, and a brief discussion of each model's outcomes is given in \citet{Lammer2021}.\\

One key feature that all of the models share is that the terrestrial planets experienced a protracted history of late accretion subsequent to planet formation. After core formation and the initial separation of the silicate reservoirs (e.g. the crust and mantle), leftover planetesimals on planet-crossing orbits were consumed by the terrestrial bodies as mass supplements \citep{Wetherill1977,Day2012}. This took the form of impact bombardment by comets \citep{Gomes2005}, planetesimals remaining after the primary phase of accretion \citep{Bottke2007}, and asteroids from the main belt \citep{MintonMalhotra2010,Nesvorny2017a} and the purported E-belt \citep{Bottke2012}. This late accretion thermally, structurally, and chemically modified solid surfaces, most of which is directly visible as craters today. \\

In the current Solar System, small objects that venture near the Earth -- so called near-Earth objects -- consist of asteroids and comets; the latter are mostly the Jupiter-family comets. The asteroids that venture near the Earth -- the near-Earth asteroids -- are typically injected onto Earth-crossing orbits from the main asteroid belt through the $\nu_6$ secular resonance near 2~au, the 3:1 Jovian mean-motion resonance near 2.5~au, and the Mars-crossing population \citep[e.g.][]{Gladman1997,MorbidelliGladman1998,Bottke2002,Granvik2018}. Main belt asteroids are driven towards Mars-crossing orbits by numerous weak mean-motion, secular, and three-body resonances \citep{Migliorini1998,NesvornyMorbidelli1998,MorbidelliNesvorny1999}, as well as the Yarkovsky effect \citep{Farinella1998,FV1999,VF1998,VF1999,VF2000}. From there they are rapidly driven to Earth-crossing orbits, and their population declines with a half life ranging from $\sim 2.3$~Myr for objects in the $\nu_6$ and 3:1 resonances \citep{Gladman1997} to 60~Myr for the low-inclination Mars-crossers, to over 100~Myr for the high-inclination Mars-crossers \citep{Migliorini1998,Michel2000,Bottke2002,FernandezHelal2023}. \\

\begin{figure*}[ht]
\centering
 \resizebox{\hsize}{!}{\includegraphics{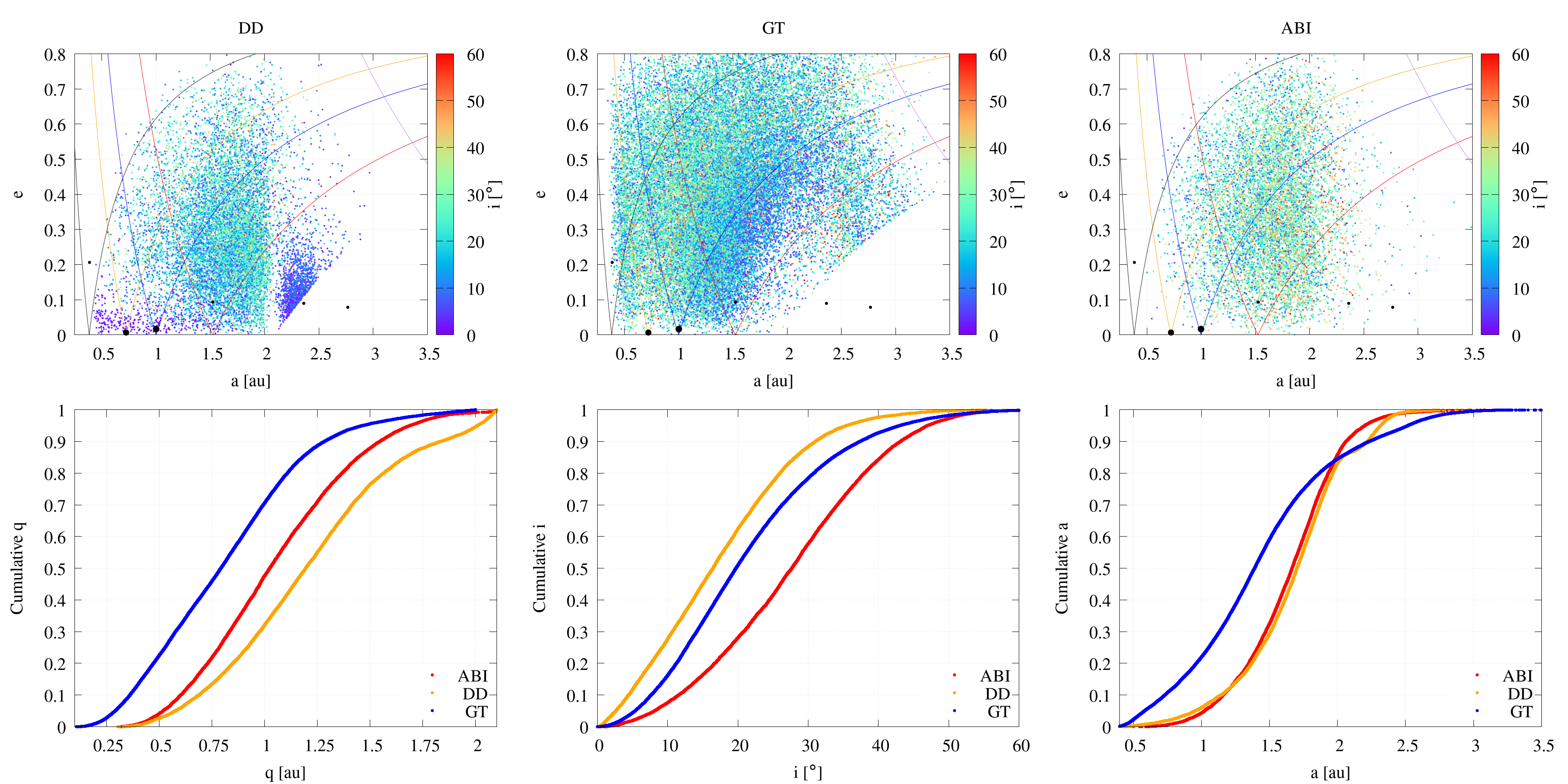}}
\caption{\label{fig:tpinit}Initial conditions of leftover planetesimals for the various models. Top row: semi-major axis versus eccentricity. The grey, orange, blue, and red lines denote the eccentricity required to cross the orbits of Mercury, Venus, Earth, and Mars, respectively. The colour coding is a proxy for the inclination, indicated by the colour bar. Bottom row: cumulative distributions in perihelion, inclination, and semi-major axis. Blue is for the grand tack mode, orange for depleted disc, and red for implantation.}
\end{figure*}

It is not obvious, however, whether in the early Solar System these same timescales modulated the impact rate on the terrestrial planets. During this early epoch the impact rate was dominated by planetesimals left over from terrestrial planet formation, whose orbital distribution was very different from that of the main belt today \citep[e.g.][]{Morbidelli2018,Brasser2020,Nesvorny2022}. A constraint on the impact rate comes from crater chronology on the Moon and Mars, which for the first 1 Gyr is fitted well by a simple exponential decline with an e-folding time of 145-150~Myr for the Moon \citep{Neukum2001,Werner2023} and 200~Myr for Mars \citep{WernerOdy2014,Werner2019}. Even though the e-folding time of the Neukum and Werner chronologies are similar, the absolute crater density of the Werner chronology is much lower, resulting in the Werner chronology predicting ages of up to 200~Myr older; this slower decline requires a much less massive impacting population. Dynamical simulations of leftover planetesimals, however, yield decline rates that vary greatly and are not necessarily consistent with that of the published crater chronologies \citep{Morbidelli2018,Brasser2020,Nesvorny2022}.\\

The reason for these discrepancies could be that the initial conditions of the planetesimals at the time of the Moon-forming event were not consistent with the population that produced the current craters on the Moon and Mars. Both \citet{Morbidelli2018} and \citet{Brasser2020} relied on leftover planetesimals from terrestrial planet simulations in the framework of the grand tack (GT) model \citep{Walsh2011}, while \citet{Nesvorny2022} relied on results from the annulus model \citep{Hansen2009}. Yet none of these initial conditions appear to reproduce the cratering rate advocated by the established crater chronologies with a good degree of satisfaction, and many different models of terrestrial planet formation are present in the literature. These models likely produce different leftover populations.\\

In this work I determine the impact rate of leftover planetesimals of three models of terrestrial planet formation simulations in my database, and whether these are or can be made consistent with modern crater chronologies. I also place constraints on what the orbital distribution of this population might have been, which of these models best reproduces this distribution, and what these constraints imply for models of terrestrial planet formation.

\section{Numerical simulations: initial conditions and methods}
\label{sec:methods}
\begin{figure}
\centering
 \resizebox{\hsize}{!}{\includegraphics{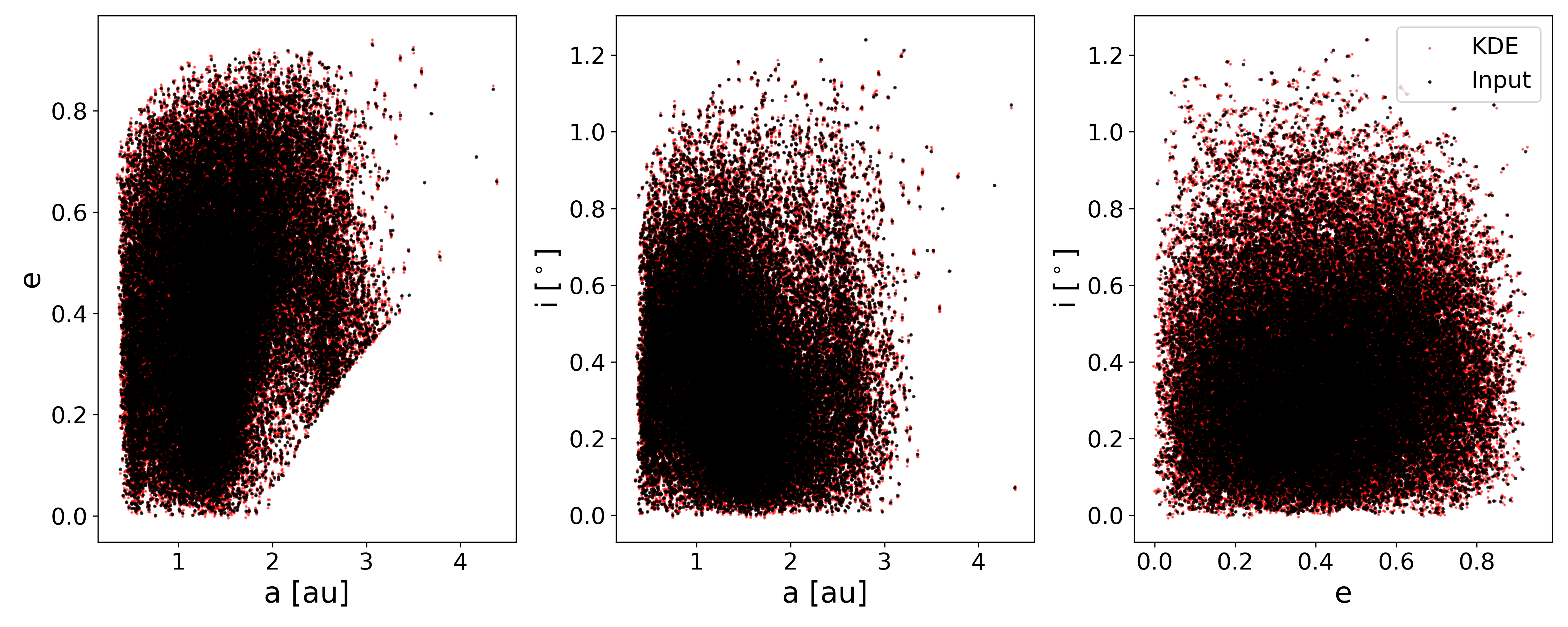}}
\caption{\label{fig:lainit}Initial conditions and clones using KernelDensity sampling for the GT model. The black dots show the initial distribution, and the red dots are the clones. }
\end{figure}

\begin{figure*}
\centering
 \resizebox{\hsize}{!}{\includegraphics{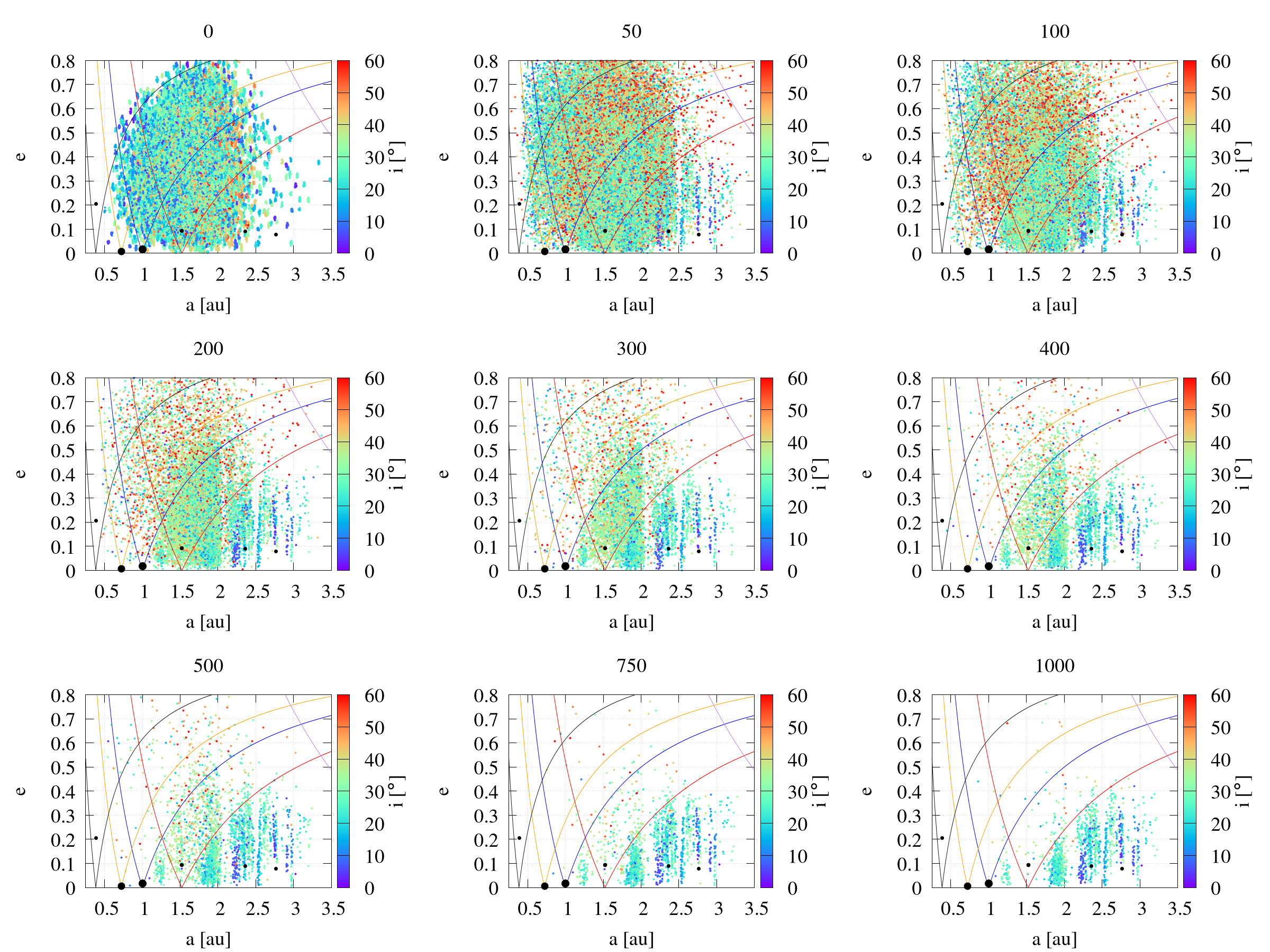}}
\caption{\label{fig:tpevo}Snapshots of the dynamical evolution of leftover planetesimals for the ABI model. The panels are for different times (in Myr), as indicated by the titles. Each panel depicts semi-major axis versus eccentricity, with the colours depicting the inclination, as indicated by the colourbar.}
\end{figure*}

This work relies on long-term simulations of massless test particles left over from simulations of terrestrial planet formation. Following \citet{Brasser2020}, the initial conditions for this work were taken from a database of terrestrial planet formation simulations in my possession based on three models: the GT model \citep{Walsh2011,Brasser2016}, the (asteroid belt) implantation (ABI) model \citep{Brasser2025}, and the depleted disc (DD) model \citep{Izidoro2014,MahBrasser2021}. Briefly, the GT model relies on the inward-outward migration of Jupiter and Saturn to truncate the planetesimal disc near 1 au to keep Mars' mass low. The DD model artificially lowers the surface density of solids beyond 1-1.5 au to starve the region near Mars of material to accrete. The ABI model relies on the implantation of planetesimals near the gas giants to account for terrestrial accretion of carbonaceous chondrite-like material \citep[e.g.][]{Dauphas2017}. I took a snapshot of the position and velocity vectors of the planetesimal population 60 Myr after the start of the simulations, which approximately coincides with 4.5 Ga. From these data I filtered out all planetesimals that had a perihelion distance $q<2$~au (2.1~au for DD to establish how the inner main belt affects the decline), and an aphelion distance $Q<4.5$~au so that they do not immediately venture near Jupiter. The initial values of semi-major axis ($a$), eccentricity ($e$) and inclination ($i$) are shown in the top three panels of Fig.~\ref{fig:tpinit}. The bottom three panels show the cumulative distributions of perihelion, inclination, and semi-major axis for all three models. All cumulative distributions are similar to either logistic or (offset) Rayleigh distributions. Specifically, the cumulative perihelion and inclination distributions are (offset) Rayleigh, and the semi-major axis distributions are logistic. These distributions are given by
\begin{eqnarray}
    F(x|\,\mu,\,\sigma) &=& 1-{\rm e}^{-(x-\mu)^2/2\sigma^2}\: {\rm (Rayleigh)}, \\
    F(x|\,\mu,\,\sigma) &=& (1+{\rm e}^{-(x-\mu)/\sigma})^{-1}\: {\rm (logistic)},
\end{eqnarray}
where e is the base of the natural logarithm. The fitting constants for the perihelion are $\sigma_q=0.55-0.65$~au, with the widest for DD, and with $\mu_q$ equal to the minimum $q$ of about 0.35~au for ABI, 0.41 and DD and 0.1 au for GT. For the inclination I obtain $\sigma_i = 14^\circ-23^\circ$ with $\mu_i=0$. For the semi-major axis $\mu_a = 1.40$~au for GT and 1.65~au for DD and ABI, and $\sigma_a=0.20$~au for ABI and DD, and 0.32~au for GT. Uncertainties in the fit are $<2^\circ$ for the inclination and $<0.02$~au for $q$ and $a$.\\

The GT model initially has the most Earth-crossers (70\% of all planetesimals), while the DD model has the least (only 30\%). The ABI model has the highest average inclination, while this is lowest for the DD model. The planetesimals in the GT model are generally closer to the Sun than in the ABI and DD models, for which the cumulative semi-major axis distributions are very similar up to 2~au. Given these different initial conditions the expectation is that the dynamical evolution of all three populations will evolve somewhat differently.\\

\begin{figure}
\includegraphics[height=0.31\textheight,width=0.5\textwidth]{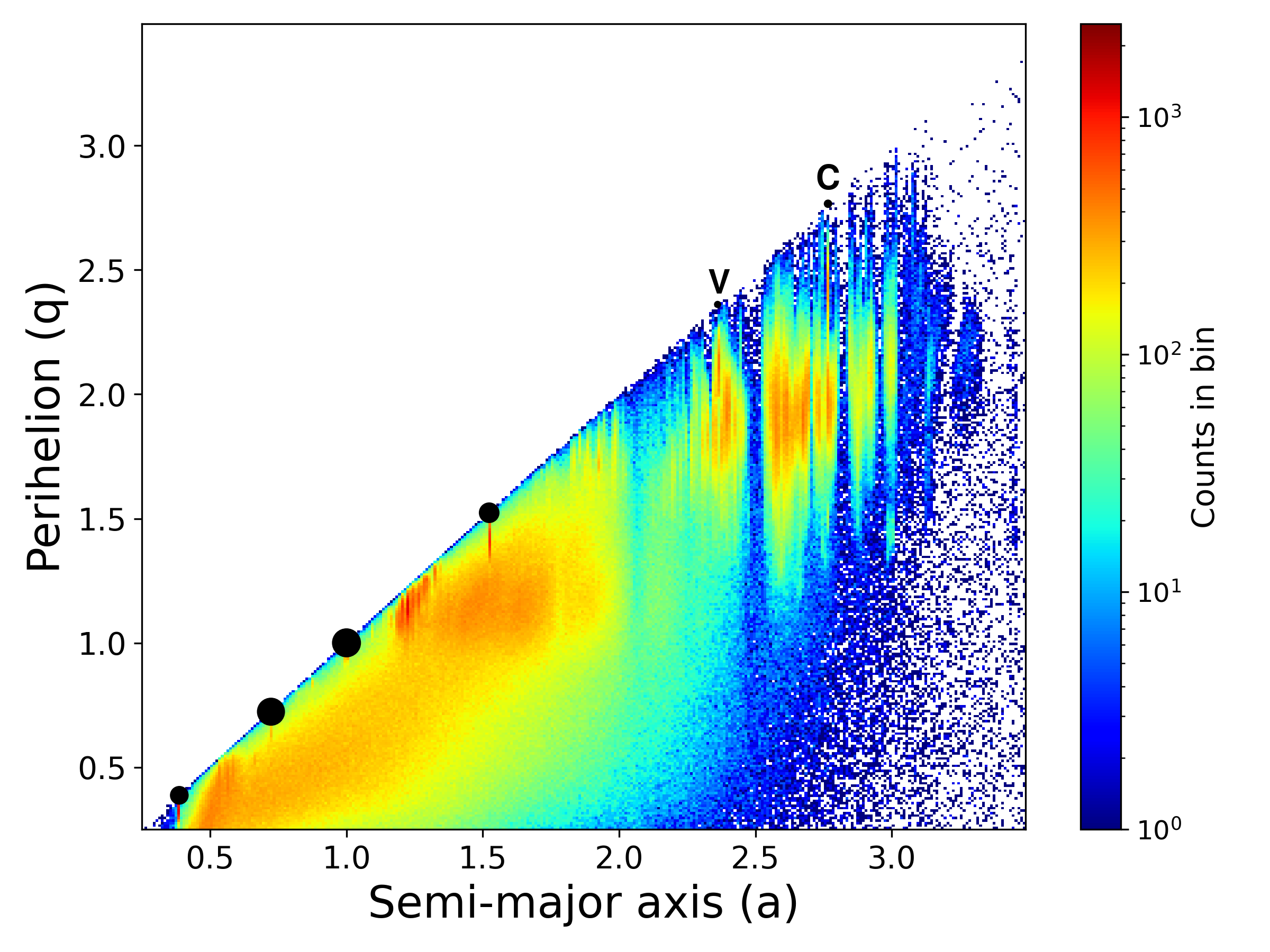}\\
\includegraphics[height=0.31\textheight,width=0.5\textwidth]{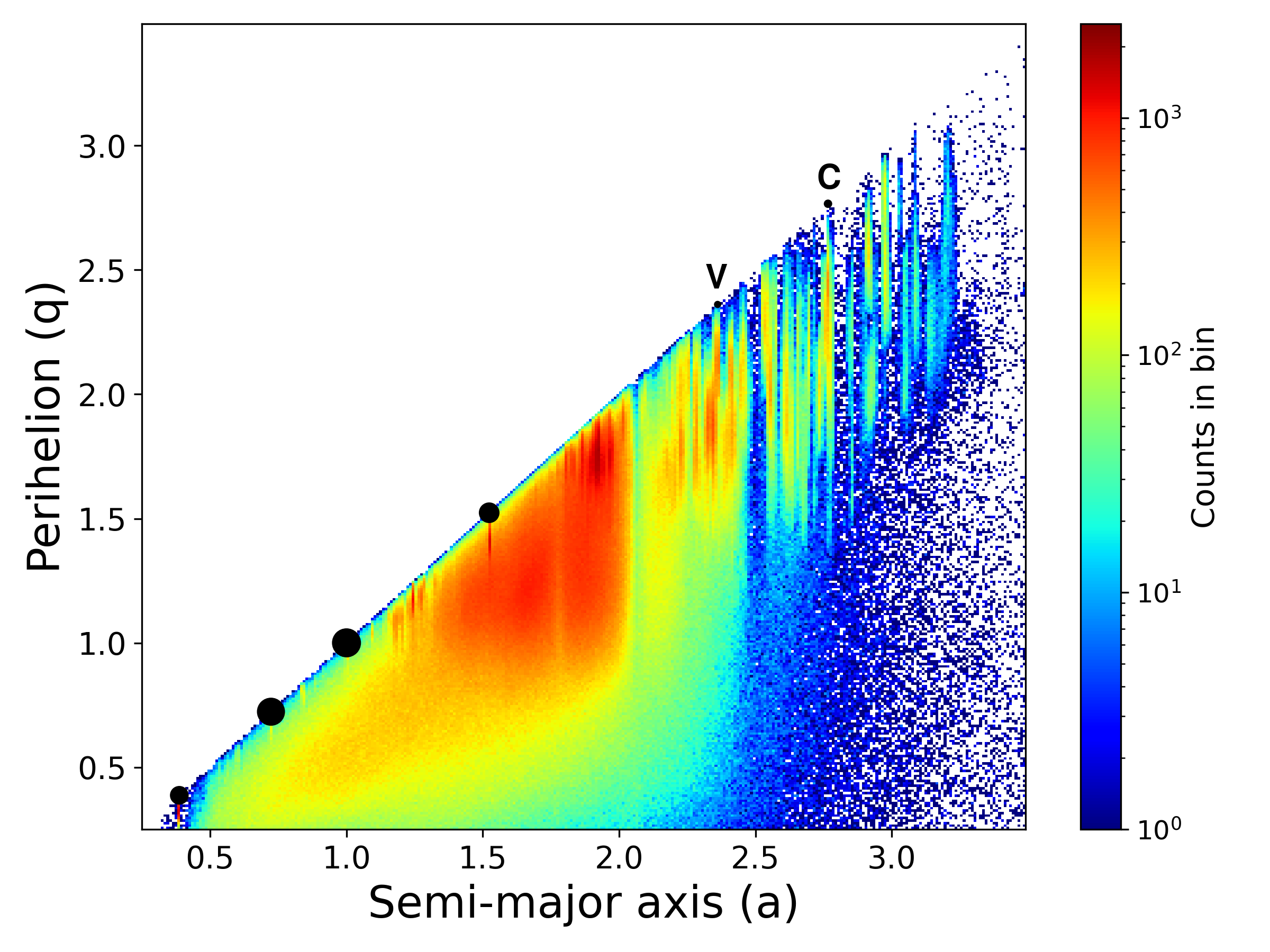}\\
\includegraphics[height=0.31\textheight,width=0.5\textwidth]{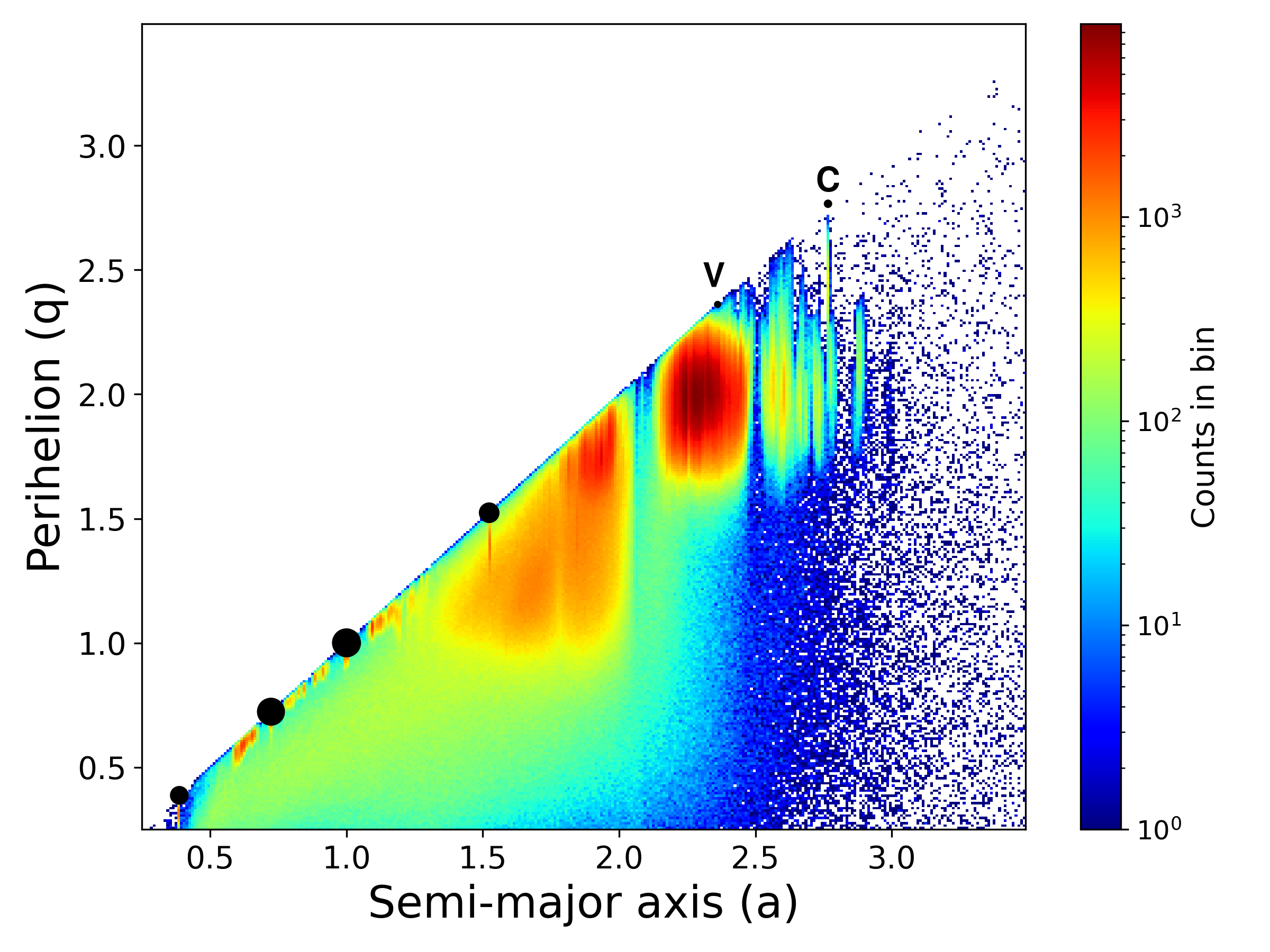}\\
 \caption{\label{fig:heat}Heat maps of the dynamical evolution of planetesimals showing the semi-major axis versus perihelion for all three models: GT (top), ABI (middle), and DD (bottom). The density increases from blue to red, with blue dots indicating few planetesimals ventured in a specific region of phase space, and red indicating that many remained in said region. The planets are indicated by large black bullets. The letters V and C stand for Vesta and Ceres.}
\end{figure}

The initial number of planetesimals from each set of terrestrial planet formation simulations was between 9\,000 (9k) and 35\,000 (35k), so I created clones of the initial test particle distribution using the KernelDensity Python module from the scikit-learn package \citep{Pedregosa2012}. A Gaussian kernel with a width of 0.005 was used on the semi-major axis, eccentricity and inclination; the other three angles were chosen uniformly random between 0 and 360$^\circ$. An example of this cloning in shown in Fig.~\ref{fig:lainit}. The width parameter was obtained through trial and error: I experimented with a larger width parameter, but I found that by increasing the width the eccentricity distribution of the clones, which was the most sensitive to the kernel width, became too wide.\\

The dynamical simulations were run using the central processing unit (CPU) version of the Gravitational ENcounters with GPU Acceleration (GENGA) N-body integrator \citep{GrimmStadel2014,Grimm2022}. All simulations were run on the Vega supercomputer in Slovenia using four CPU threads per simulation. All the major planets were added on their current orbits: Mercury to Neptune, plus dwarf planets Vesta and Ceres. I realise that the orbits and masses of the current planets are different from those that existed in the planet formation simulations; this procedure gravitationally shocks the orbits of the test particles. None of the terrestrial planet formation simulations, however, reproduced the terrestrial planets exactly as they are today, and I am not aware of an alternative approach for generating the initial conditions.\\

Simulations were run for 1 Gyr with a time step of 0.01~yr ($10^{11}$ time steps in total). Each simulation began with 8192 test particles representing the leftover planetesimal population. Simulation data were written to disc every 100\,000 years (100 kyr). Planetesimals were removed once they were further than 40 au from the Sun (whether bound or unbound), or when they collided with a planet or when they were closer than 0.05 au from the Sun. To keep simulation time reasonable I did not include corrections from General Relativity because this results in a 33\% speed penalty. The code takes about 3-6 million steps per particle per second on Vega using four threads on AMD EPYC 7H12 CPUs clocked at 3.1 GHz; maximum speed was achieved when the number of test particles $N \geq 128$. I used two CPU threads when $64\leq N<128$ and only one when $N<64$. For each terrestrial planet formation model I ran 16 simulations with a total of 131k test particles.\\

\section{Results}
\subsection{Global dynamical evolution}
In Fig.~\ref{fig:tpevo} I show snapshots of the semi-major axis versus eccentricity for the ABI simulations. After 400~Myr almost all Earth and Venus crossers are eliminated, but there is still a substantial population of Mars-crossers. There is also a small group of planetesimals that are seemingly stable between the Earth and Mars for 1 Gyr. This is an artefact of the simulations because these bodies would have had their orbits modified by the Yarkovsky effect and therefore this region is expected to be empty -- as it is today. Simulations of these same initial conditions including the Yarkovsky effect are ongoing and will be discussed in future works. Last, there is a long-term stable population at moderate inclination between Mars and the $\nu_6$ resonance, which is located at 2.05 au at low inclination \citep{WilliamsFaulkner1981}; this population is akin to the Hungaria family \citep{Bottke2012}. \citet{MorbidelliNesvorny1999} showed that bodies beyond the $\nu_6$ resonance can slowly leak onto Mars-crossing orbits due to chaotic motion induced by resonance overlap. In addition, some bodies began and remained on long-term meta-stable orbits in the main belt with semi-major axis between 2.1 and 3 au and perihelion distance $q> 1.8$~au. \\

\begin{figure}
\centering
 \resizebox{\hsize}{!}{\includegraphics{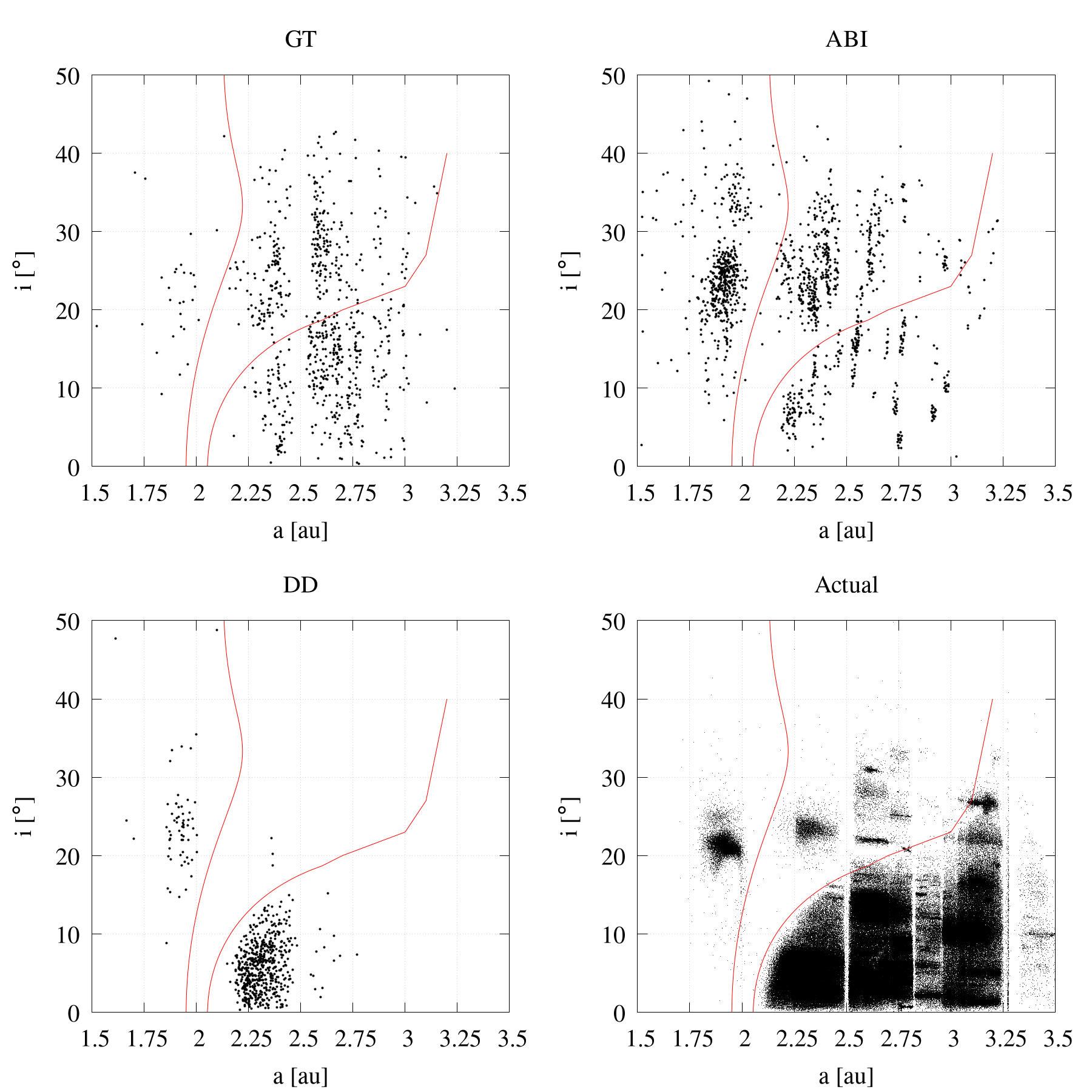}}
\caption{\label{fig:tpab}Snapshots at 1 Gyr of the planetesimals showing the semi-major axis versus inclination compared to the actual main belt. The red lines are the $\nu_{16}$ and $\nu_6$ secular resonances. Panel titles indicate the model used. Asteroid belt data were downloaded from AstDys-2.}
\end{figure}

A different visualisation of the evolution of the planetesimals is shown in Fig.~\ref{fig:heat}, which is a heat map of the semi-major axis and perihelion distance of all bodies in all simulations for each model. The terrestrial planets plus Vesta and Ceres are shown with large bullets. These heat maps show the counts per bin and not the residence time, though it is easy to convert between the two. It is immediately clear that the maps for the ABI and DD models show some similarities, while the map for the GT model is rather different. \\

The distributions shown in the heat maps depend on the initial conditions for the models. In the ABI and DD models, there is a sharp decrease in density, or residence time, for planetesimals that are Earth-crossing versus those that are not Earth-crossing, as indicated by the two red triangular regions in these panels. In both the ABI and DD models there is one such region with $q>1$~au and $1.2<a<2$~au, while in the DD model the second one has $q>1.5$~au and $2<a<2.5$~au. The second region is in the DD model. It is more heavily populated than in the same regions in the ABI and GT models due to the higher initial perihelion cutoff and the higher implantation for the DD model. The boundaries in semi-major axis for the second region are marked by the $\nu_6$ resonance at 2 au and the 3:1 Jovian resonance at 2.5~au. In both plots the E-belt region between 1.7 au and 2 au, as proposed by \citet{Bottke2012}, remains occupied for a long time. In the DD model the inner main belt is heavily populated, a consequence of the initial conditions. In contrast, the GT heat map looks substantially different, with the reddest region having $1<q<1.3$~au and $1.2<a<2$~au apart from a small sample with $a\sim 1.2$~au and $e\lesssim 0.03$. The GT model has the greatest proportion of initial Earth-crossers, so that it is expected that this population has a shorter average lifetime than those of the ABI and DD models. The GT heat map also has the orange region extend all the way down to Mercury, while in the ABI and DD models it stops at Venus. The reason for this difference is that the initial perihelion distribution for the GT model extends all the way down to 0.1~au while this is 0.3~au for the ABI and DD models. One immediate outcome of this is that the impact probability with Mercury in the GT model is a factor of three higher than for the other two models. The heat maps suggest that the initial perihelion distribution plays an important role in the rate of decline of the population, and thus in the impact rate onto the inner planets.\\

In Fig.~\ref{fig:tpab} I present a snapshot of the semi-major axis-inclination distribution at the end of the simulations for all three different cases, and compare them to the actual main belt in the bottom right panel. Even though the initial conditions used here all came from terrestrial planet formation models, all of these models leave some material in the main belt region. Comparing its orbital distribution with that of the present main belt serves as a constraint of the validity of those models, and associated predictions for the long-term bombardment of the terrestrial planets from the main belt. It is clear that the DD case fares poorly in that it only produces the Hungaria group near $a\sim 1.8$~au and $i \sim 20^\circ$ and the inner main belt underneath the $\nu_6$ secular resonance for which $2.1<a<2.5$~au, but not the high inclination population above the $\nu_6$. The GT and ABI cases populate all of the main groups to some extent, except the one between 3~au and 3.25~au. The ABI model leaves more mass in the Hungaria group than the GT case, while GT populates the middle main belt between 2.5~au and 2.8~au better than ABI does. Overall the GT and ABI models fare better than DD.

\subsection{Impact chronology and population decline}
\begin{table}
    \caption{Impact fitting constants for Earth and Mars, and the fraction remaining.}
    \begin{tabular}{l|llll}
     Planet & $\alpha_1,\tau_1$ & $\alpha_2,\tau_2$ & $\alpha_3,\tau_3$ & $\alpha_4,\tau_4$\\ \hline
     Earth GT &0.13, 9.3& 0.69, 34.2 &0.16, 89.5& 0.012, 303\\
     Earth ABI &0.068, 8.0 & 0.46, 36.8& 0.46, 109&0.016, 351 \\
     Earth DD &0.10, 6.4& 0.43, 30.8&0.44, 105& 0.021, 306 \\ \hline 
     Mars GT & 0.12, 11.9 & 0.65, 46.4 & - & 0.23, 154 \\
     Mars ABI & - & 0.20, 32.2 & 0.75, 116 & 0.044, 272 \\
     Mars DD & 0.16, 16.8 & - & 0.64, 78.5 & 0.21, 206 \\ \hline
     Rem GT &0.17, 8.5 & 0.56, 35.1 & 0.23, 88.6 & 0.016, 1470 \\
     Rem ABI  &0.13, 10.9 & 0.41, 42.0& 0.45, 116& 0.023, 1370  \\
     Rem DD & 0.20, 17.9 &0.43, 65.0 & 0.27, 143 & 0.10, 5233
    \end{tabular}
    \tablefoot{Uncertainties in the fitting constants are $<10\%$. Time unit is Myr. `Rem' stands for = `remain'.} 
    \label{tab:chronfit}
\end{table}

The complementary cumulative distribution of impacts of planetesimals onto the Earth and Mars, as well as the fraction of remaining planetesimals is fitted as
\begin{equation}
    F(>t) = \sum_{i=1}^N \alpha_i\, {\rm e}^{-t/\tau_i},
\end{equation}
where the constants $\alpha_i$ are weights adding up to 1, the variables $\tau_i$ are e-folding times, and e is the base of the natural logarithm; the number of terms required turned out to be usually four and sometimes three. I prefer this formulation rather than using a stretched exponential -- $\ln F(>t) = -(t/\tau)^\beta$ -- because the multi-exponential formulation yields a better fit, and the fitting constants yield insights into the initial conditions and fundamental dynamical processes; specifically, the number of terms reveals the number of sinks or processes involved. The results of the fits are given in Table~\ref{tab:chronfit}. For impacts on the Earth the values of $\tau_2$ and $\tau_3$ are rather similar across the models, being identical to within 25\% and having values near 35~Myr and 100~Myr; the difference increase to to 37\% for $\tau_1$.  \\

\citet{Nesvorny2017a} used a similar fit with three terms to model the impact rate rather than the cumulative number of impacts. The first time constant fitted by \citet{Nesvorny2017a} is 37~Myr and the second is 160~Myr. Both are of the same order as mine. \citet{Nesvorny2017a} do not provide an explanation for these time constants and just state they are required to fit their high impact flux in the first Gyr. I argue that the clustering of ages around specific values for different models suggests that these time constants reflect intrinsic processes in the system, and that the fitting constants $\alpha_i$ are a reflection of the initial orbital distribution of the planetesimals.\\

For impacts on Mars, the middle two time constants in the ABI model are similar to those for the Earth, but they carry different weights; the martian impact time constants are more different from Earth's for the other two models. I found that a fit with three terms generally worked better than one with four terms. The same equation also works well for the decline of the whole population, although the last term has an e-folding time that is much longer than the simulation time. I find that the exponential fit at large $t$, which corresponds to the term with $\tau_4$, is only an approximation: experimentation shows that here a linear fit would also suffice, akin to the constant cratering flux assumption \citep{Neukum1975}. In particular, for the GT model the first three time constants fitting the decline of the whole population are similar to impacts on Earth, while the weights are identical within factors of 25\%. For impacts on Mars the time constants of its impact chronology for the DD model resembles the overall decline of the whole population. This indicates that, to first order, the impact rate is modulated by the overall rate of decline, or removal, of the planetesimals from the reservoirs.\\

These results beg the question whether the coefficients $\alpha_i$ are related to the initial fraction of Earth (and Mars) crossers. In the GT, ABI, and DD models the initial fraction of Earth-crossers is 70\%, 50\% and 35\%, which are close to the coefficients $\alpha_2$ for impacts on Earth. For the ABI and DD models $\alpha_2 \sim \alpha_3$ for Earth, so the remaining particles not controlled by Earth and only by Mars impact the Earth on a timescale $\tau_3$. For impacts on Mars only three terms are needed, and for the ABI and DD models martian impacts occur mostly on timescales $\tau_3$ and $\tau_4$, in contrast to $\tau_2$ for the GT model. For all models the impact rate on the terrestrial planets is related to the overall decline of the whole population As such, if no planetesimals remain to impact Mars because most of them are gone after a few times $\tau_2$, then the rate of martian impacts must also decline in tandem.\\

In summary, the time constants for both the impacts and the remaining fraction seem to be clustered in four groups: one group at $9 \pm 3$~Myr, another at $35 \pm 10$~Myr, a third at $95 \pm 15$~Myr and a fourth that is generally $>200$~Myr. This clustering suggests that the rate of impacts is partially set by the rate of decline of planetesimals, which in turn is set by their leakage rate from their initial reservoirs. For comparison: \citet{Nesvorny2017a} used three terms and found $\tau_2=37$~Myr, $\tau_3=160$~Myr and $\tau_4=1500$~Myr for the rate of impacts on the Earth, even though their study only included the main asteroid belt plus E-belt subjected to an episode of giant planet migration. These similar values support my earlier hypothesis that the time constants are features of the system. Furthermore, the crater chronologies of \citet{Neukum2001} and \citet{Werner2023} have e-folding timescales of $\tau_{\rm NW}\sim 150$~Myr, which is of the same order as $\tau_3$, indicating that the lunar craters mostly reflect the impacts from an ancient Mars-crossing population. This result implies that the number of Earth-crossers at the time that the Moon formed had to be low, probably less than half that of the Mars-crossers.\\

\begin{figure*}
\includegraphics[height=0.31\textheight,width=1.0\textwidth]{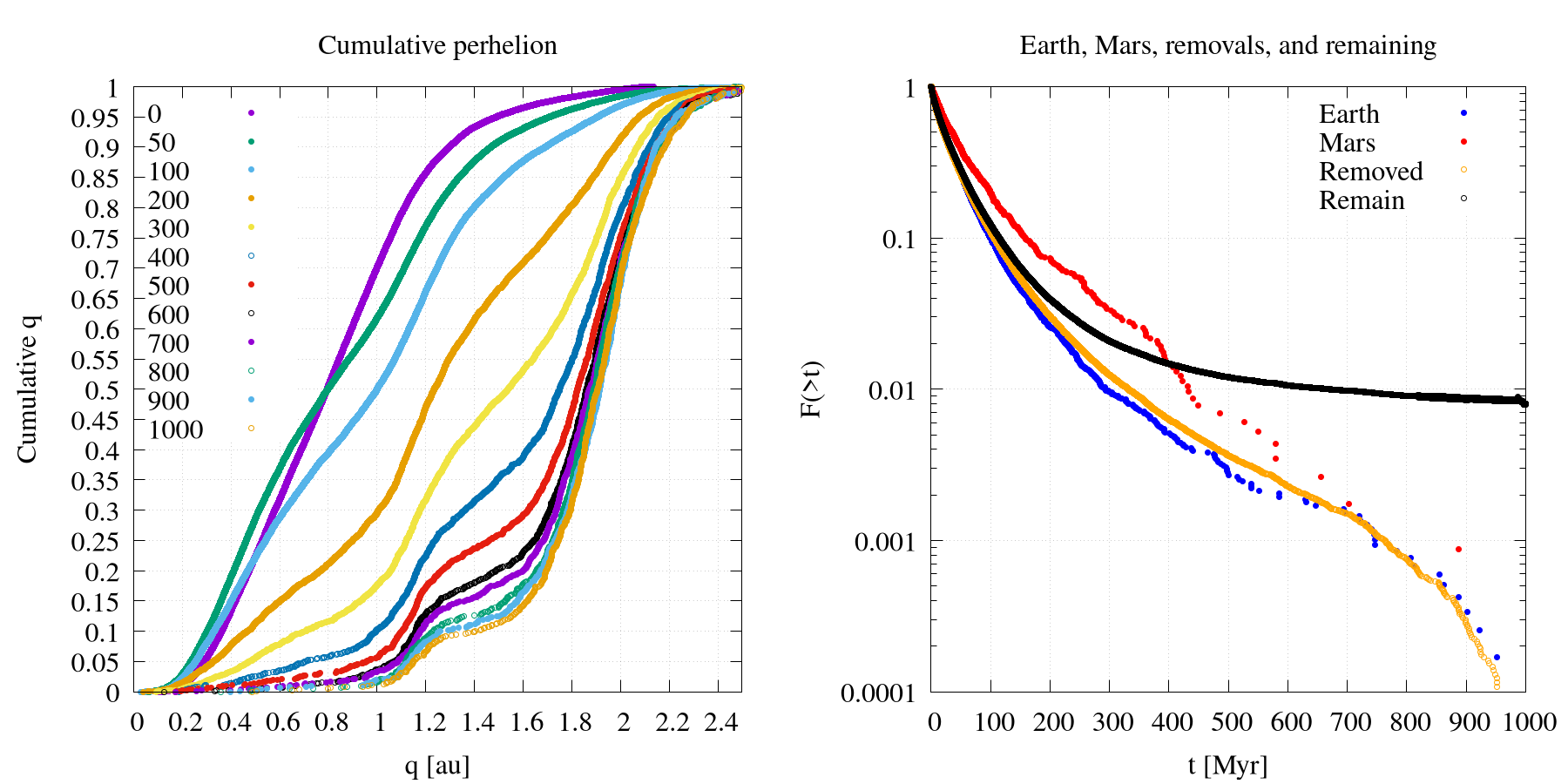}\\
\includegraphics[height=0.31\textheight,width=1.0\textwidth]{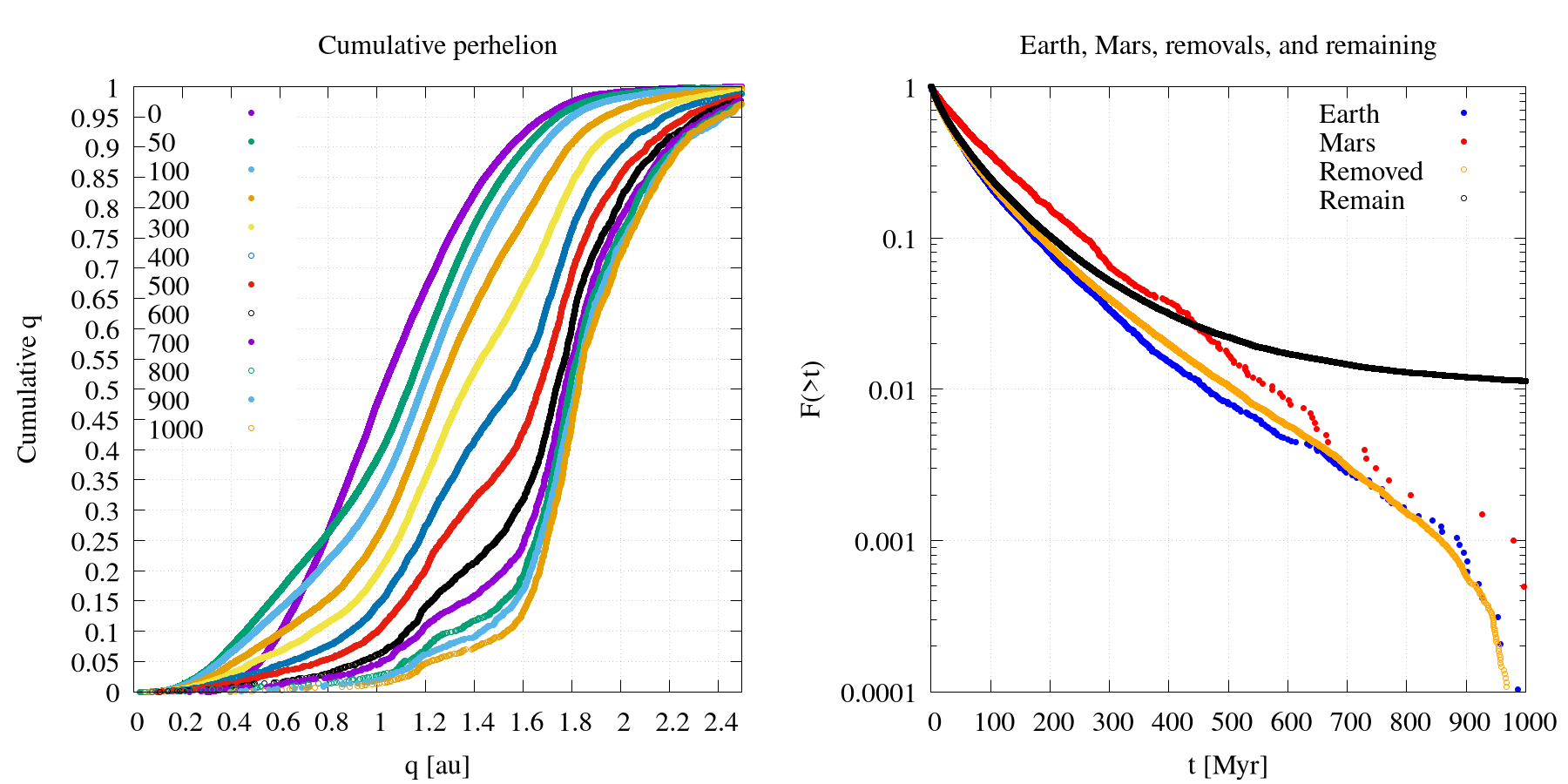}\\
\includegraphics[height=0.31\textheight,width=1.0\textwidth]{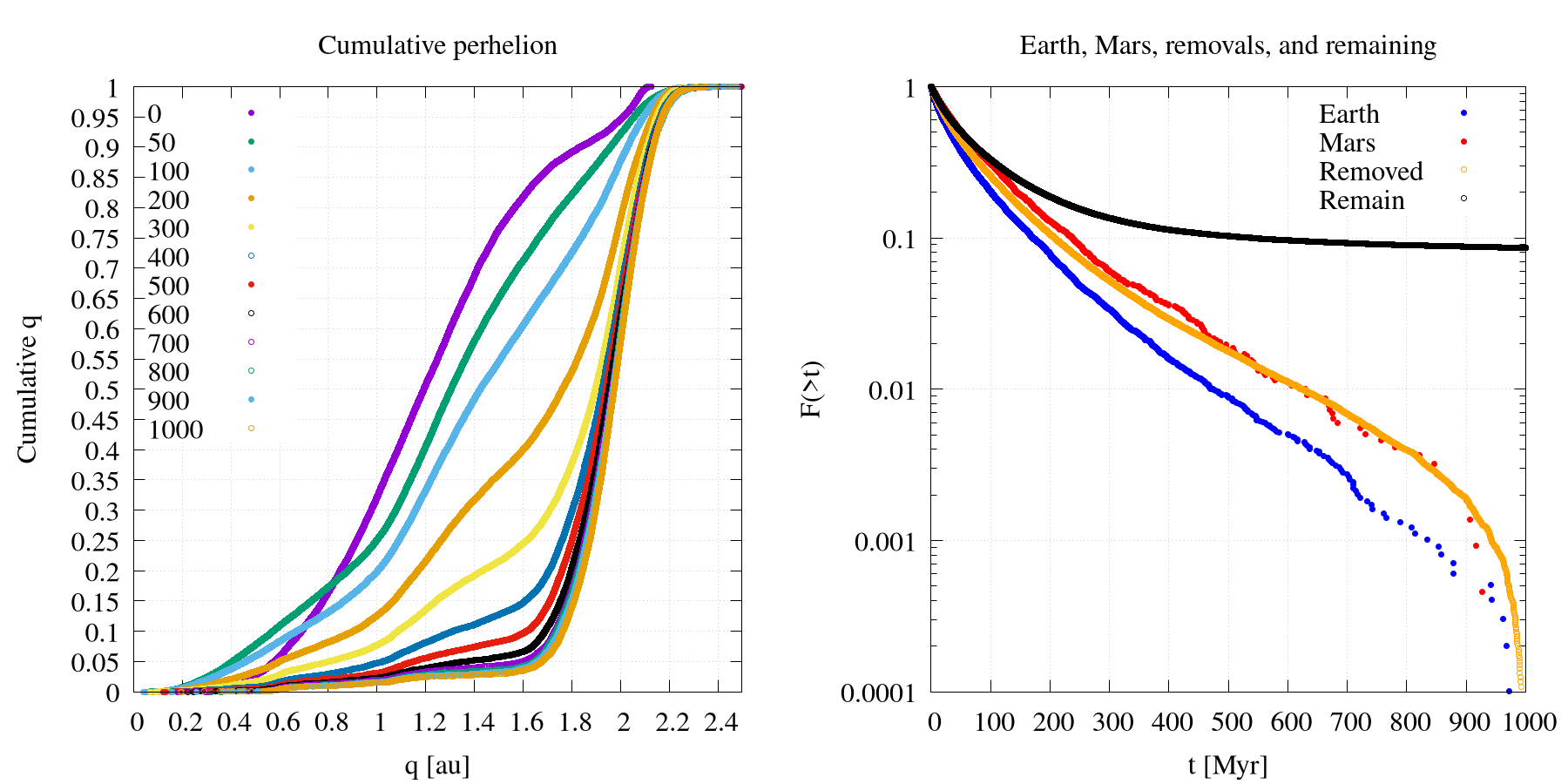}\\
 \caption{\label{fig:qdist}Left column: Evolution of the cumulative perihelion distribution. Right column: Impacts on Earth (blue) and Mars (red), the fraction of planetesimals remaining (black), and the total planetesimals that are removed (orange)  for GT (top), ABI (middle), and DD (bottom).}
\end{figure*}
\begin{figure*}
\includegraphics[height=0.48\textheight,width=1.0\textwidth]{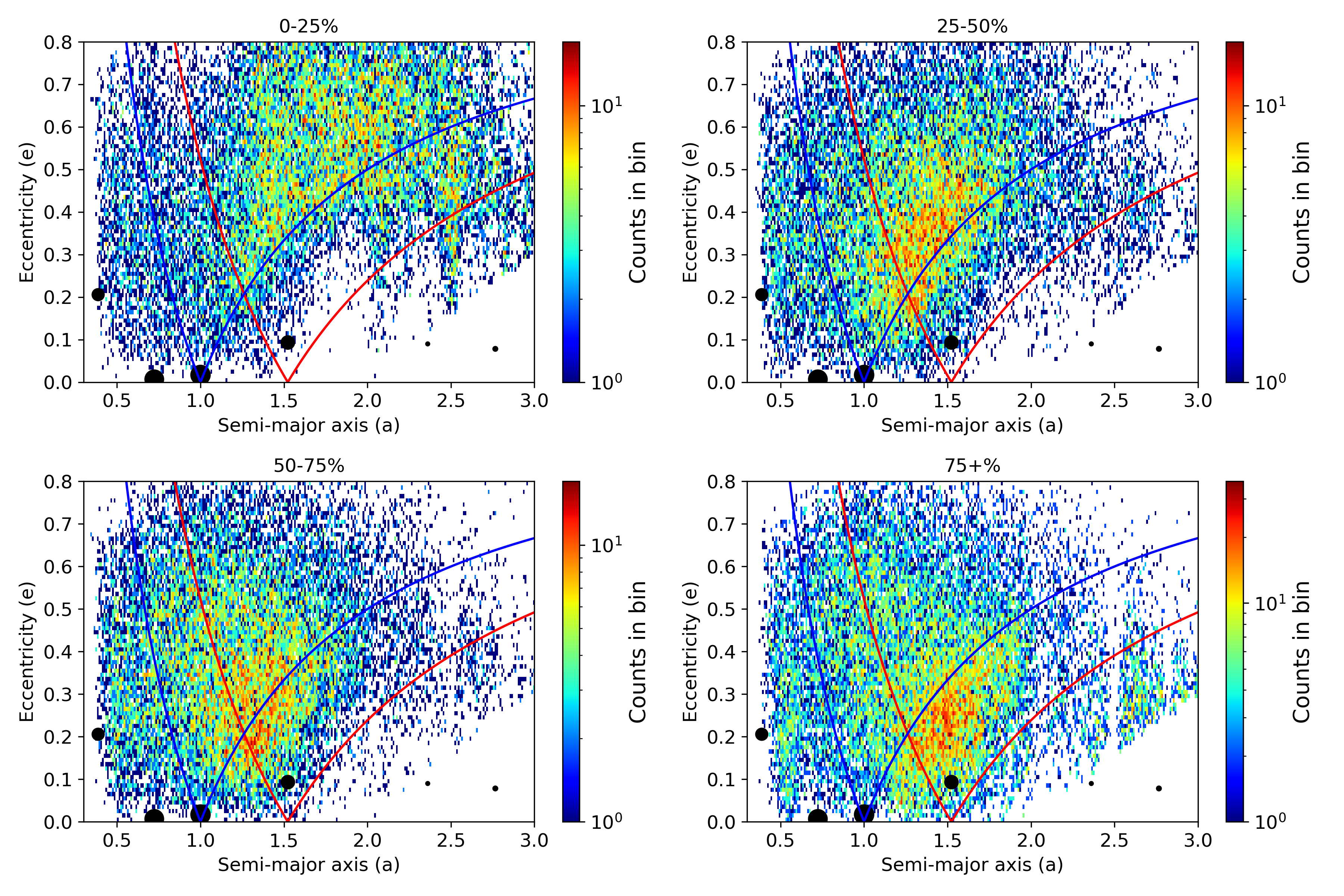}
\hrule
\includegraphics[height=0.48\textheight,width=1.0\textwidth]{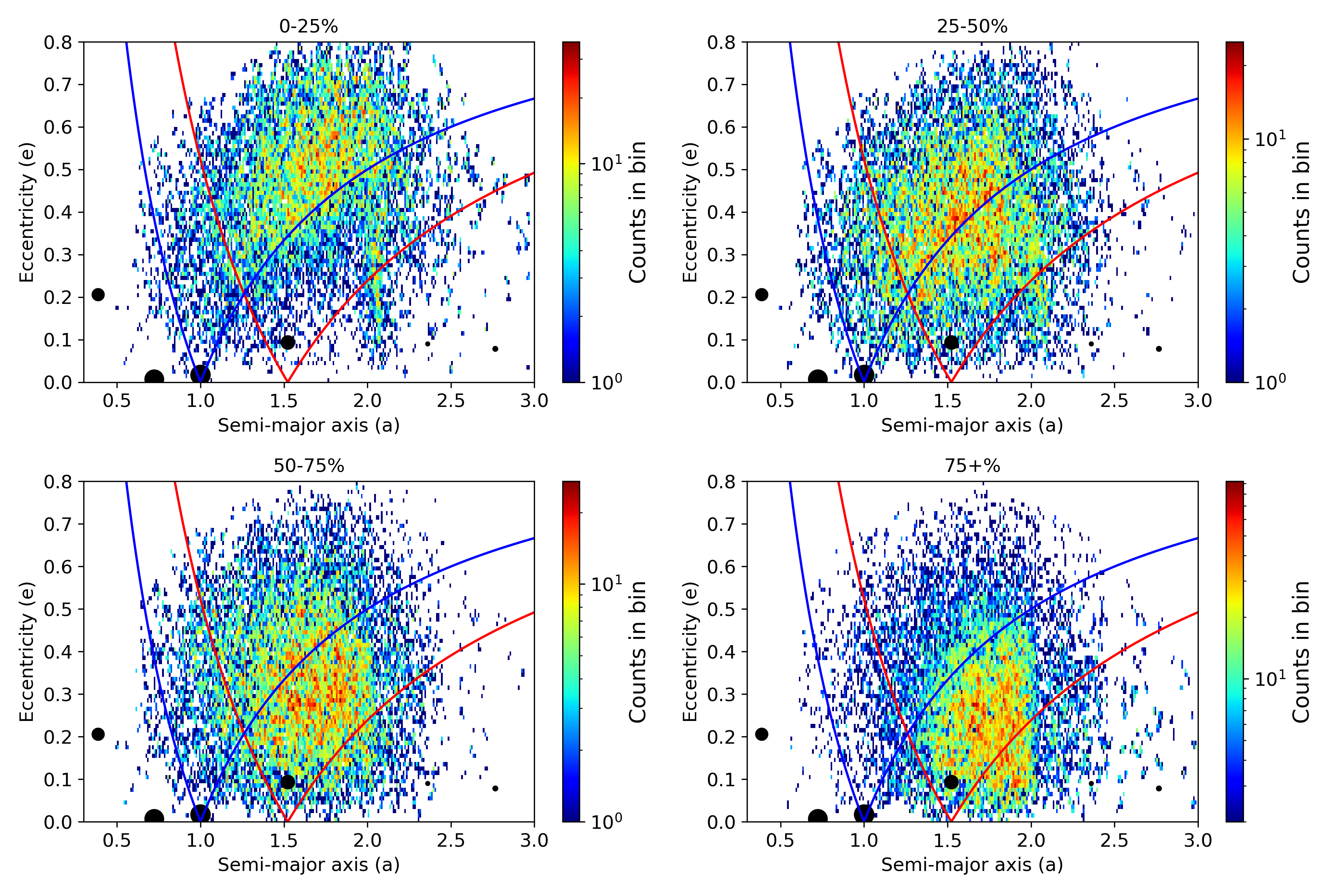}
\caption{\label{fig:hdecay}Heat maps of the initial semi-major axis versus eccentricity of planetesimals that are removed at different quartiles during the simulation. Red and blue lines indicate Mars- and Earth-crossing orbits. The top four panels are for the GT model and the bottom four panels are for the ABI model. Panel titles indicate the quartile.}
\end{figure*}

\subsection{The evolution of the perihelion distribution}
In Fig.~\ref{fig:qdist} I plot the evolution of the cumulative perihelion distribution for each model and the complementary cumulative distributions of the impacts on Earth and Mars, as well as the fraction of planetesimals remaining and the total planetesimals that are removed. In all models the cumulative $q$ distribution is normalised to the number of remaining planetesimals and rapidly changes shape, with the fraction of Earth-crossers dropping quickly, and the fraction of Mars-crossers reducing more slowly. At late stages, usually after 700~Myr, almost all of the remaining planetesimals have $q>1.6$~au. The rate of change of the cumulative $q$ distribution at specific perihelion values can be approximately quantified as follows. \\

I denote by $F(<q)$ the fraction of remaining planetesimals that is still Earth or Mars crossing. In other words, $F(<q)$ corresponds to the cumulative value of the perihelion distribution plotted in Fig.~\ref{fig:qdist}. From the simulation output I conclude that for all three models the fraction of Earth-crossers -- $F(q<1)$ -- decreases exponentially with an e-folding time of $\tau_q=230$~Myr. For example, after 500~Myr in the GT model the fraction of Earth-crossers $F(q<1)\sim 0.05$ while initially it is 0.7. The approximate e-folding time of the decline of the remaining population that is still Earth crossing is then $(500~{\rm Myr})/\ln(0.7/0.05) \sim 190$~Myr. For the fraction of Mars-crossers -- $F(q<1.5)$ -- I find an e-folding time of 220~Myr in the DD model, while the e-folding time is increased to 430 Myr for the GT and ABI models. Thus, for the DD model the value of $F(q<1)$ declines at the same rate as $F(q<1.5)$, but in the ABI and GT models the fraction of planetesimals that are still Mars crossing declines much more slowly. The rate of decline of $F(<q$) for the Earth and Mars-crossers is generally much lower than the decline in impacts and that of the whole population.\\

I argued that the initial perihelion distribution is the main constraint in determining the impact chronology on the Earth and Mars as well as global population decline. This argument is evident from the figure: the initial population decline and impact rate for the GT model is much more rapid than for the ABI and DD models, which is evident from the fitting constants $\alpha_i$ in Table~\ref{tab:chronfit}. For the ABI and GT models, the terrestrial impacts track the global decline for the first 100~Myr and begin to deviate from there, with the normalised rate of impacts on Mars being always lower than on Earth. In contrast, the terrestrial impact chronology closely tracks all the planetesimals depletion in the GT and ABI models, while Mars' impact chronology is initially off to a slower start but its impact rate converges after $\sim$200 Myr. Beyond this point, in all models the impact chronology is set by the dynamical evolution of the Mars-crossers, which evolve on a timescale of about 80--100~Myr \citep{Michel2000}, in other words $\tau_3$, because encounters with Mars will need to supply the planetesimals to Earth, either directly or via the $\nu_6$. \\

The behaviour of the remaining fraction of planetesimals is different: after a few hundred million years of evolution it levels off due to planetesimals being parked on long-lived orbits beyond Mars. Some of these slowly leak onto Mars-crossing orbits due to chaotic diffusion \citep{MorbidelliNesvorny1999}, and some of these planetesimals collide with the terrestrial planets. 

\subsection{Removal mechanism behind the fitted timescales}
To gain a greater insight into the origin of the timescales, I investigated which planetesimals are removed at what time, and from where. Figure~\ref{fig:hdecay} is a heat map showing the original semi-major axis and eccentricity of all planetesimals that are removed from the simulation at different time quartiles. The top left panel shows the removal for the first quartile of the population, the top right panel the second quartile, the bottom left shows the third quartile, and the bottom right shows the last quartile. The top four squares are for the GT model, and the bottom four are for the ABI model. Throughout the simulation planetesimals are removed from all over, as evidenced by the blue dots, and instead I focus on the orange regions. \\

For the GT model in the top left panel, which lasts for 7~Myr and is comparable to $\tau_1$, removed planetesimals initially reside on Earth-crossing orbits with semi-major axis between 1.2 and 2.5 au. There are clear indicators of rapid removal of planetesimals beginning near 2 au and 2.5 au even at low initial eccentricity, corresponding to the $\nu_6$ and 3:1. The next quartile, shown in the top right panel, occurs between 7~Myr and 21~Myr and still has the densest part on Earth-crossing orbits but at shorter semi-major axis, between 1.1~au and 1.6~au. The high density region shifts to shorter semi-major axis in the third panel, lasting from 21~Myr to 52~Myr. Combined, these quartiles last for about 1.5$\tau_2$. The reason for this shift towards shorter semi-major axis is that it takes longer to eliminate these objects through collision or by scattering them into the $\nu_6$ or the 3:1. As such, it appears likely that $\tau_2$ is related to the rate of removal of Earth-crossers. The last panel, lasting 948~Myr, shows the most slowly declining population, which are Mars-crossers with semi-major axes between 1.5 au and 2 au. This last quartile is dominated by $\tau_3$ and $\tau_4$ -- although the latter has a low coefficient $\alpha_4$ -- which suggests that $\tau_3$ is the timescale of evolution of Mars-crossers. \\

For the ABI model the top left panel has the highest density of planetesimals initially on Earth-crossing orbits with initial semi-major axes between 1.3~au and 2~au, with also some bodies in the $\nu_6$ -- see the vertical points near 2 au at low eccentricity. This quartile lasts for 12~Myr, comparable to $\tau_1$, suggesting that $\tau_1$ is related to removal through the $\nu_6$ and high-eccentricity Earth (and Venus) crossers. The next quartile, lasting from 12~Myr to 36~Myr, still has a large number of Earth-crossers, but there are also a fair amount of Mars-crossers in there; the $\nu_6$ region is still being emptied out. By now half of the planetesimals are removed. During the third quartile, lasting from 36~Myr to 91~Myr, most of the planetesimals that are removed are Mars-crossers with initial semi-major axes between 1.2 au and 2 au. The rate of population decline begins to decrease, with the terms with $\tau_2=42$~Myr and $\tau_3=116$~Myr setting the rate, once again suggesting that $\tau_2$ is related to the removal of Earth-crossers and $\tau_3$ to that of Mars-crossers. The fourth quartile lasts for 900~Myr, is modulated by the terms with $\tau_3$ and $\tau_4$ at late stages, and consists mostly of Mars-crossers. However, the E-belt region beyond Mars, with initial semi-major axis between 1.6 au and 2 au and perihelion $q>1.6$~au \citep{Bottke2012}, also plays a role; this region appears to play almost no role in the GT simulations. \\

In summary, I argue that $\tau_1$ is likely a measure of the decline of high-eccentricity Earth-crossers and bodies initially in the $\nu_6$ and 3:1, $\tau_2$ describes Earth and Venus crossers, and $\tau_3$ pertains to the Mars-crossers, and possibly the E-belt. The term $\tau_4$, which is the last few percent, is due to the decline of bodies originally beyond Mars on more stable orbits such as the inner main belt.\\

One way to verify the mechanisms behind $\tau_2$ and $\tau_3$ is to use the equations developed by \citet{Opik1951}. The probability per orbital revolution of coming within a distance $s$ of the planet is given by

\begin{equation}
    p=\frac{s^2 U}{\pi \sin i |U_x|},
\end{equation}
where $U$ is the encounter velocity scaled by the planet's orbital velocity, and $U_x$ is the $x$-component of this velocity -- which is essentially the radial component of the encounter velocity. For the initial conditions considered in the simulations the value of $U/[\pi \sin i |U_x|]$ for the Earth ranges from 1.5 to 2.2, and $\langle U \rangle \sim 0.65$. The radius $s$ at which a collision occurs can be computed from energy and angular momentum conservation, and is given by $s=R_p(1+2m_p/[M_\odot R_pU^2])^{1/2}$; when $U=0.65$ I compute that $s\sim 1.16R_p$ for the Earth and 1.13$R_p$ for Venus. This leads to a probability of collision with the Earth or Venus of $p=(4-5)\times10^{-9}$ per orbit of a planetesimal that crosses the Earth's and/or Venus's orbit. The corresponding e-folding time for collision is then $T=a^{3/2}/p \sim 400$~Myr using a typical semi-major axis of $a \sim 1.5$~au. This timescale is far longer than the timescales for removal numerically computed here. In the simulations only $\sim 20\%$ of planetesimals are removed through collision with an inner planet; the rest are removed by `ejection', that is, they are being scattered into the $\nu_6$ secular resonance or 3:1 mean-motion resonance through repeated encounters with the inner planets. These resonances pump up the eccentricities of the planetesimals and they ultimately collide with the Sun or encounter Jupiter \citep{Gladman1997}. \\

To calculate the removal timescale I need an estimate of the cumulative change in semi-major axis due to scattering. I wrote a Fortran program that implements the Monte Carlo method of \citet{Arnold1965} using \"{O}pik's equations as shown in \citet{Valsecchi1997,Valsecchi2003} for encounters with the Earth and Venus;  the addition of Mars is trivial but was not considered. Sinks include collisions with Venus and Earth, a perihelion beyond 4.5~au, entrance into the $\nu_6$ secular resonance, and ejection ($e>1)$. I considered only encounters within 60~$R_p$. The initial semi-major axis, perihelion distance and inclination were drawn from the Rayleigh and logistic distributions discussed in Sect.~\ref{sec:methods}. The mean dynamical lifetime is $68.4^{+176}_{-67}$~Myr (2$\sigma$) and the median is 36.1~Myr. The lifetime distribution is roughly logistic in $\log t$, with fitting parameters $\log \mu_t=1.53$ (corresponding to $\mu_t=34$~Myr), and width $\log \sigma_t = 0.74$. It is known that \"{O}pik's method does not adequately describe the decline of planetesimals \citep{Dones1999}, so I consider these estimates to be upper limits. In any case, a rough estimate of the average dynamical lifetime of a planetesimal through repeated Earth and Venus encounters is $\sim$70~Myr, which is about 2$\tau_2$.  \\

\begin{figure}
\centering
 \resizebox{\hsize}{!}{\includegraphics{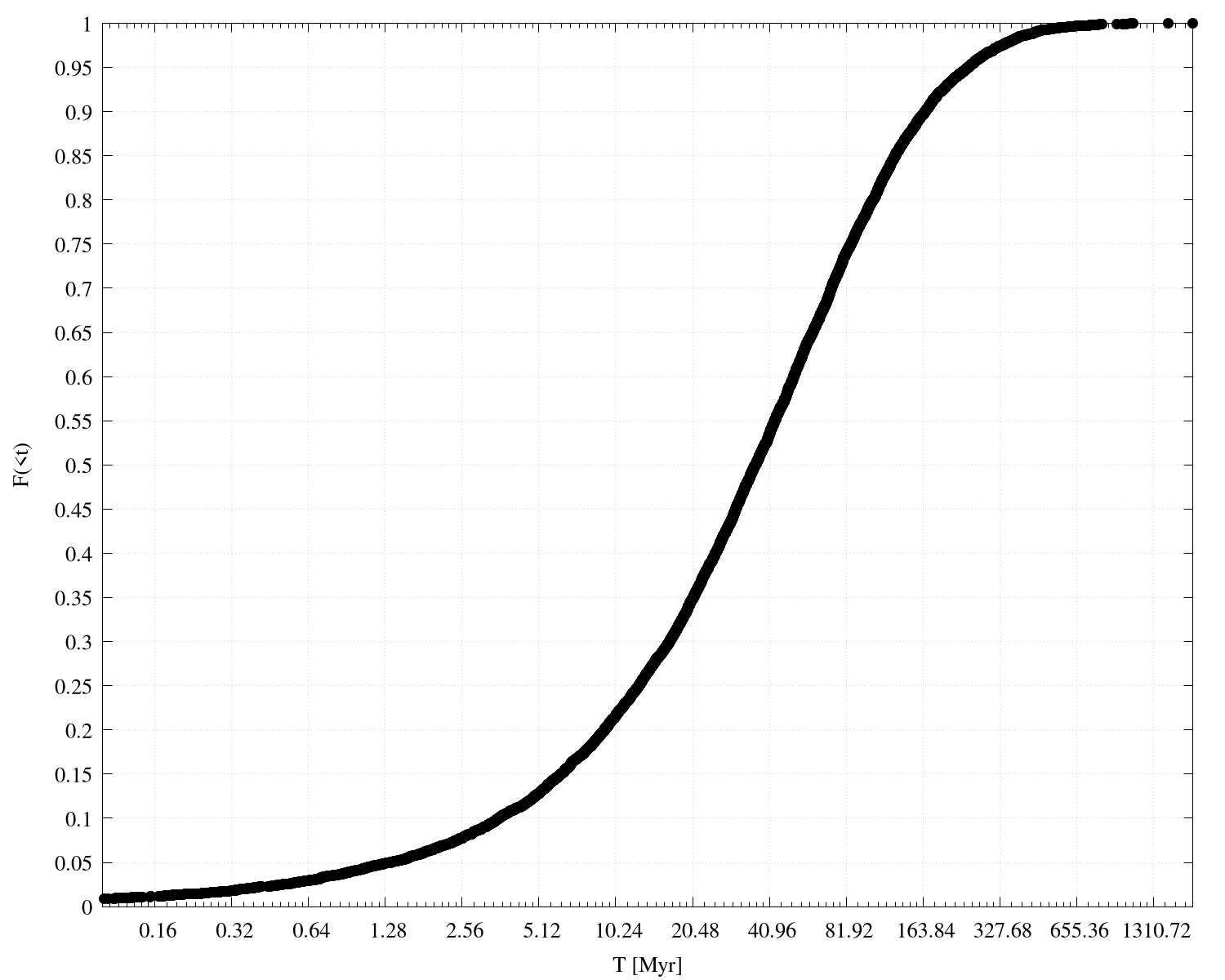}}
\caption{\label{fig:tlife}Cumulative distribution of the dynamical lifetime of planetesimals scattered by the Earth and Venus using the Monte Carlo method of \citet{Arnold1965}, which relies on \"{O}pik's equations. The median lifetime is close to the value of $\tau_2$ found from numerical simulations.}
\end{figure}

For Mars $\langle U \rangle \sim 0.6$ and $s=1.03R_p$ so that $p=1.2\times10^{-9}$. Elimination through encounters and collisions with Mars results in an e-folding time of about 200~Myr when $a=1.5$~au, which is twice as long as found numerically here as well as in the literature \citep{Michel2000,FernandezHelal2023}, but still of the same order. Objects near Mars undergo large changes in eccentricity due to the proximity of the $\nu_6$ secular resonance; this can send them onto Earth-crossing orbits, which in turn leads to their rapid removal. Still, I argue that $\tau_3$ is associated with the removal of planetesimals by interaction with Mars.\\

For both models the rate of decline during the last quartile, which lasts $900\pm 50$~Myr for all models, is dominated by Mars-crossers, while during the second and third quartile the rate is still influenced by remaining Earth-crossers. However, for the GT model $\tau_3=89$~Myr for the decline of the population, akin to the timescale found by \citet{Michel2000} for the Mars-crossers. Given the similarity between the initial cumulative semi-major axis distribution in the ABI and DD models and the lower impact rate of the latter, it stands to reason that the initial perihelion distribution is the most important parameter controlling the impact rate. The fitting constants $\alpha_i$ therefore probably represent the (initial) perihelion and semi-major axis distribution. \\

One final point of note pertains to the total fraction remaining after 1~Gyr. For both the GT and ABI models this is about 1\%, but for the DD model this is about 10\%. The last value is problematic because the current mass in the main belt is about $6\times 10^{-4}$~$M_\oplus$, which probably declined by about 50\%-65\% over the past 4~Gyr \citep{MintonMalhotra2010,Nesvorny2017a,Brasser2020}, with even less than that in the inner main belt. In the DD model the remaining 10\% resides in the inner main belt, implying that the initial population in planetesimals at 4.5~Ga was no greater than $10^{-3}$~$M_\oplus$. This is insufficient to account for the lunar craters and the abundances of highly siderophile elements in the lunar mantle \citep{Morbidelli2018,Brasser2020}. Therefore, given the simulations that I have run here, I will have to cautiously discount the initial conditions of the DD model as being viable for late accretion.

\section{Spherule layer constraints on the Archean impact rate and source population}
It has been suggested that terrestrial spherule beds are a record of ancient terrestrial impacts \citep{SimonsonGlass2004}. \citet{JohnsonMelosh2012} used the spherule layer thickness from the recorded list of known spherule beds in \citet{GlassSimonson2012} to compute the diameter range of the impactor that created the spherule bed. It is not clear whether this list is complete, but I will treat it as such. In Table~\ref{tab:jm} I have listed the names and ages of the largest of these spherule beds, the average impactor diameter from the range listed in \citet{JohnsonMelosh2012}, and the corresponding number of impacts on the Earth for objects with diameter $D_{\rm i}>1$~km using a size-frequency distribution slope of $-2$. For completeness I added the Sudbury layer with the proposed impactor diameter using the formula from \citet{JohnsonMelosh2012}, but I do not use it in the analysis. There is also the Vredefort crater in South Africa with an approximate age of 2 Ga, which may have left traces of a bed as far as Russia \citep{Huber2014,Allen2022}. No spherule beds have been found between 0.6 and 1.7~Ga ago, but this time interval has not been extensively explored for impact spherules \citep{SimonsonGlass2004}.\\

\citet{Bottke2012} used the presence of the spherule layers and their thickness to compute the impact rate onto the Earth from a declining population of asteroids placed just beyond the orbit of Mars. Here I will use a slightly different approach. From Table~\ref{tab:jm} I compute that there were 17k impacts on Earth between 3.47~Ga and 2~Ga of planetesimals with diameter $D_{\rm i}>1$~km. This results in an average impact rate of 11.5~Myr$^{-1}$. The uncertainties are factors of a few due to the possible range of the impactor diameters. What reservoir provided these planetesimals? \\

\begin{table}
    \centering
    \caption{Archean impactor spherule beds.}
    \begin{tabular}{l|ccc}
 Name & Age [Ga] & $D_{\rm i}$ [km] & $N(D_{\rm i}>1)$\\ \hline
 Sudbury & 1.82 & 15 & 225 \\
  Gr\ae nses\o  & 2.0 & 60 & 3600 \\
  Kuruman  & 2.46 & 15 & 225 \\
  Dales George  & 2.49 & 40 & 1600 \\
  Bee George  & 2.54 & 14 & 196 \\
  Reivilio  & 2.56 & 22 & 484 \\
  Paraburdoo  & 2.57 & 22 & 484 \\
  Monteville  & 2.62 & 38 & 1444 \\
  S4  & 3.24 & 43 & 1849 \\
  S3  & 3.25 & 56 & 3136 \\
  S2  & 3.26 & 48 & 2304 \\
  S1  & 3.47 & 41 & 1681
         \end{tabular}
    \tablefoot{The diameters of the impactors are the average from \citet{JohnsonMelosh2012}. Two other layers, Jeerinah and Carawine, may be from the same impact as Monteville \citep{GlassSimonson2012}.} 
    \label{tab:jm}
\end{table}
During the Archean and later, the cratering rate on the Earth can be computed analogous to the outer Solar System, and is simply $\dot{C}=P_{\rm E}F_{\rm EC}$ \citep{BWW2025}, where $F_{\rm EC}$ is the flux of objects on Earth-crossing orbits, and $P_{\rm E}$ is the impact probability with the Earth. The results in Fig.~\ref{fig:qdist} seem to apply that this relation is justified by the similar shape of the terrestrial and martian impact curves with those of removed planetesimals. Furthermore, at late times ($>500$~Myr of simulation time) most objects are entirely beyond Mars so that the impact rate is determined by the rate at which these objects become Mars-crossers. This situation is similar to that of Kuiper belt and scattered disc objects that leak into the realm of the giant planets from beyond Neptune \citep{DuncanLevison1997}.\\ 

In practice, the flux of objects leaking in from beyond Mars is equal to $F_{\rm EC}=|r|N$ \citep{Duncan1995}, where $N$ is the total number of objects in all reservoirs with $D_{\rm i}>1$~km. I followed \citet{Duncan1995} and computed the rate of decline of planetesimals as the number of planetesimals that left the system ($\Delta N$) divided by the final number of the particles at the end ($N_{\rm fin}$), and the time interval ($\Delta t$). In other words,

\begin{equation}
        r = \frac{1}{N_{\rm fin}}\frac{\Delta N} {\Delta t} \approx \frac{d\ln N}{dt}
    \label{eq:rsd}
.\end{equation} 
It is worth mentioning that generally $r$ depends on the time range over which it is calculated because the rate of decline is generally not constant with time; instead, the rate of decline is a function of time itself: it decreases with increasing time. An exception is an exponential decay of the form e$^{-t/\tau}$, for which we have $|r|=\tau^{-1}$, which is constant. From the simulation output at late times $|r|=\tau_4^{-1}$. In addition, apart from the small fraction of objects that are ejected due to encounters with Jupiter, the far majority are lost due to collisions with a planet or the Sun; on the way to the Sun they are temporary Earth-crossers, so that to a good approximation $F_{\rm EC}$ is equal to the rate of decline of the total remaining population (see also Fig.~\ref{fig:qdist}). However, care should be taken in using $\dot{C}=P_{\rm E}|r|N$. One could argue that the relation for impacts on Earth and the population decline are not identical, violating the relation, but that is only true if one assumes that $P_{\rm E}$ is constant with time. This is generally not the case: $P_{\rm E}$ grows with time as the population declines and planetesimals are eliminated. Only at late stages, when the planetesimals are leaking in from beyond Mars, is the assumption that $P_{\rm E}$ is constant approximately valid.\\

In the simulations, the impact probability with Earth ranges between 7.3\% (ABI and DD) and 9\% (GT), so I will take 7.5\% as the average. This impact probability implies that 227k objects with $D_{\rm i}>1$~km were lost between 3.47~Ga and 2~Ga from the various small body reservoirs in the inner Solar System. Assuming that $|r|=\tau_4^{-1}$ and $\tau_4 \sim 1.5$~Gyr then the aforementioned 227k objects represent a decline of 63\% by 2~Ga in the population remained at 3.47~Ga. A similar value of $\tau_4=1.5$~Gyr was also found by \citet{Nesvorny2017a}. Using $\tau_4=1.5$~Gyr implies that there were about 617k objects with $D_{\rm i}>1$~km at 3.47~Ga available to supply the terrestrial impactors. The main belt between 2.1~au and 3.3~au barely declines \citep{MintonMalhotra2010,Nesvorny2017a, Brasser2020}, so that these 617k planetesimals must have been in the leftover population, or in the E-belt, or both.\\

The fit to impacts on Earth and Mars suggest that $\tau_4 \sim 300$~Myr rather than $\sim 1.5$~Gyr from the fit to the remaining planetesimals. This would suggest a discrepancy between the two fits and that choosing $\tau_4 = 1.5$~Gyr is incorrect. However, upon closer inspection, this is not the case: it can be shown by doing a Taylor expansion of the fits at large $t$ (e.g. at 800 Myr) that the rates of decline of the impacts and the whole population are equal to within a factor of two. In addition, if indeed $\tau_4=300$~Myr, then the population remaining at 3.47~Ga would have declined a further 99.3\% by 2~Ga. Such an extreme depletion can be ruled out because it would imply an implausibly large mass in the initial population at 4.5~Ga that violates constraints from the lunar cratering record and the terrestrial zircon record \citep{Brasser2021}. \\

At 3.47~Ga the remaining population is about 1\% (ABI and GT) or 9\% (DD) of the population at 4.5~Ga. The remaining fraction in the DD simulations is too high, so I will use a total decline of 99\% after 1~Gyr of evolution. This implies a primordial population of the order of 60 million objects with $D_{\rm i}>1$~km at 4.5~Ga, which corresponds to a mass in leftover planetesimals of about 0.015~$M_\oplus$. This estimate is consistent with earlier values reported in the literature \citep{Brasser2021,Nesvorny2022}. \\

What about the E-belt of \citet{Bottke2012} being the source? \citet{Brasser2020} conducted dynamical simulations of the E-belt for 1 Gyr. For this work these were extended to 2~Gyr. For the E-belt the impact probability with Earth is $P_E=5\%$, requiring 340k bodies to leak from the E-belt between 3.47~Ga and 2~Ga to supply all of the terrestrial impacts. The fraction remaining can be fitted as $F(>t)=0.66{\rm e}^{-t/133\,{\rm Myr}}+0.33{\rm e}^{-t/552\,{\rm Myr}}$, so that at late times $|r|=552$~Myr$^{-1}$. In the simulations the E-belt loses an additional 93\% of its mass between 3.47~Ga and 2~Ga, so that I can safely assume that at 3.47~Ga the E-belt had 340k objects with $D_{\rm i}>1$~km. The simulation output shows that at 3.47~Ga the E-belt only has 3.9\% of its mass at 4.5~Ga, implying that the primordial number of E-belt objects with $D_{\rm imp}>1$~km is approximately 8.7M assuming no collisional evolution. The required mass for this many objects is about 0.002~$M_\oplus$, which is a factor of a few higher than what was suggested by \citet{Bottke2012}. It is therefore more likely that the Archean impacts that created the terrestrial spherule beds were predominantly ($\sim 75\%$) supplied by leftover planetesimals on meta-stable orbits slowly leaking onto Earth-crossing orbits from beyond Mars.

\section{Conclusions}
I have simulated the dynamical evolution of leftover planetesimals for 1 billion years from three different models of terrestrial planet formation: the GT \citep{Walsh2011}, the DD \citep{Izidoro2014}, and ABI \citep{Brasser2025}. Each one of these models yields a different initial distribution of orbital elements for the leftover planetesimals. In particular, the initial fraction of Earth-crossers for the GT is twice as high as for DD, which has implications for the dynamical evolution and impact rate onto the Earth and Mars. The DD model leaves too much mass after 1~Gyr to be consistent with the current mass of the main asteroid belt.\\

I fitted the cumulative number of impacts and the remaining fraction of planetesimals with a function that is a sum of three to four exponentials whose e-folding timescales cluster around 10~Myr, 35~Myr, 100~Myr, and $>$200~Myr. These distinct timescales are the result of different processes removing the planetesimals at different rates, with high-eccentricity Earth-crossers and planetesimals in the $\nu_6$ secular resonance being removed first ($\tau_1$), followed by Earth-crossers ($\tau_2$), Mars-crossers ($\tau_3$), and finally objects leaking onto Mars-crossing orbits from the E-belt and the inner main asteroid belt ($\tau_4$). The timescale $\tau_3$ is of the same order as that of the lunar chronology of \citet{Neukum2001} and \citet{Werner2023}, suggesting that the lunar craters are sourced predominantly from an ancient population of Mars-crossers, and that the number of Earth-crossers at the time that the Moon formed had to be rather low. As such, the initial perihelion distribution of the leftover planetesimals plays a decisive role in the overall rate of decline of the population and the impact rates onto the Earth and Mars. These results place constraints on models of terrestrial planet formation, wherein the number of Earth-crossers at the time of the Moon's formation had to be lower than Mars-crossers.\\

To calibrate the impact rate at late times, I used the Archean spherule beds as listed in \citet{JohnsonMelosh2012}, converting the average diameter of the impactor to the number of objects with diameter $D_{\rm i}>1$~km striking the Earth over 1.47~Gyr between 3.47~Ga and 2~Ga. I conclude that the most likely source of these impacts are the leftover planetesimals rather than the main belt or E-belt asteroids, and that this Archean flux requires a primordial leftover population with a mass of about 0.015~$M_\oplus$ at 4.5~Ga. Implications for the lunar cratering rate and the amount of mass accreted by the Moon and Mars will be presented in a future paper.

\begin{acknowledgements}
I thank Stephanie Werner and Emily W. Wong for comments, and John Chambers for his constructive review. I acknowledge EuroHPC Joint Undertaking for awarding the project EHPC-REG-2024R01-093 access to Vega at IZUM, Slovenia. This study is supported by the Research Council of Norway through its Centres of Excellence funding scheme, project No. 332523 PHAB. GENGA can be obtained from \url{https://bitbucket.org/sigrimm/genga/}.
\end{acknowledgements}

\bibliographystyle{aa}
\bibliography{aa55873-25}
\end{document}